\documentclass[aps,prd,floatfix,reprint,twocolumn,reprint,amsmath,amssymb,superscriptaddress,showpacs]{revtex4-1}
\usepackage{graphicx}
\usepackage{bm}
\usepackage{amsmath}
\usepackage{dcolumn}
\usepackage{dsfont}
\usepackage{amssymb}
\usepackage{tabularx}
\usepackage{array}
\usepackage{float}
\usepackage{color}
\usepackage{epstopdf}
\usepackage{mathrsfs}
\usepackage[colorlinks, linkcolor=blue,anchorcolor=blue,citecolor=blue,urlcolor=blue]{hyperref}

\begin{document}
\title{Hidden Zeeman-type spin polarization in bulk crystals}

\author{Shan Guan}
\affiliation{State Key Laboratory of Superlattices and Microstructures, Institute of Semiconductors, Chinese Academy of Sciences, Beijing 100083, China}

\author{Jun-Wei Luo}
\email{jwluo@semi.ac.cn}
\affiliation{State Key Laboratory of Superlattices and Microstructures, Institute of Semiconductors, Chinese Academy of Sciences, Beijing 100083, China}
\affiliation{Center of Materials Science and Optoelectronics Engineering, University of Chinese Academy of Sciences, Beijing 100049, China}

\author{Shu-Shen Li}
\affiliation{State Key Laboratory of Superlattices and Microstructures, Institute of Semiconductors, Chinese Academy of Sciences, Beijing 100083, China}
\affiliation{Center of Materials Science and Optoelectronics Engineering, University of Chinese Academy of Sciences, Beijing 100049, China}

\author{Alex Zunger}
\affiliation{Renewable and Sustainable Energy Institute, University of Colorado, Boulder, Colorado 80309, USA}

\begin{abstract}
Exploring hidden effects that have been overlooked given the nominal global crystal symmetry but are indeed visible in solid-state materials has been a fascinating subject of research recently. Here, we introduce a novel hidden Zeeman-type spin polarization (HZSP) in nonmagnetic bulk crystals with sublattice structures. In the momentum space of these crystals, the doubly degenerate bands formed in a certain plane can exhibit a uniform spin configuration with opposite spin orientations perpendicular to this plane, whereas such degenerate states are spatially separated in a pair of real-space sectors. Interestingly, we find that HZSP can manifest itself in both centrosymmetric and non-centrosymmetric materials. We further demonstrate the important role of nonsymmorphic twofold screw-rotational symmetry played in the formation of HZSP. Moreover, two representative material examples, i.e., centrosymmetric WSe$_2$ and noncentrosymmetric BaBi$_4$O$_7$, are identified to show HZSP via first-principles calculations. Our finding thus not only opens new perspectives for hidden spin polarization research but also significantly broadens the range of materials towards spintronics applications.
\end{abstract}

\maketitle
\emph{{\color{blue} Introduction.}} Since the discovery of hidden spin polarization in centrosymmetric crystals~\cite{Zhang2014}, the past few years have witnessed a surge of interest in condensed matter systems for both the theoretical and experimental exploration of hidden physical effects~\cite{Partoens2014,Riley2014,Wu2017a,Yao2017}, which refer to that the said effect that the global crystal symmetry would seemingly forbid it indeed exists and can be observed because it arises from the local site symmetry~\cite{Yuan2019,Lin2020}. A variety of hidden effects, including intrinsic circular polarization~\cite{Liu2015}, hidden orbital polarization~\cite{Ryoo2017}, hidden Berry curvature~\cite{Cho2018}, unconventional superconductor~\cite{Liu2017,Gotlieb2018}, layer-valley coupling~\cite{Yu2020}, and the hidden anomalous Hall effect~\cite{Zhang2020}, has been uncovered~\cite{Guan2022}. These findings not only considerably enrich the material candidates for studying the target effects that have been limited to systems with certain global crystal symmetry before, but may also provide new design principles for building novel electrically tunable devices~\cite{Yuan2019,Lin2020,Guan2020a,Guan2022}.

\begin{figure}[!t]
\centerline{\includegraphics[width=0.5\textwidth]{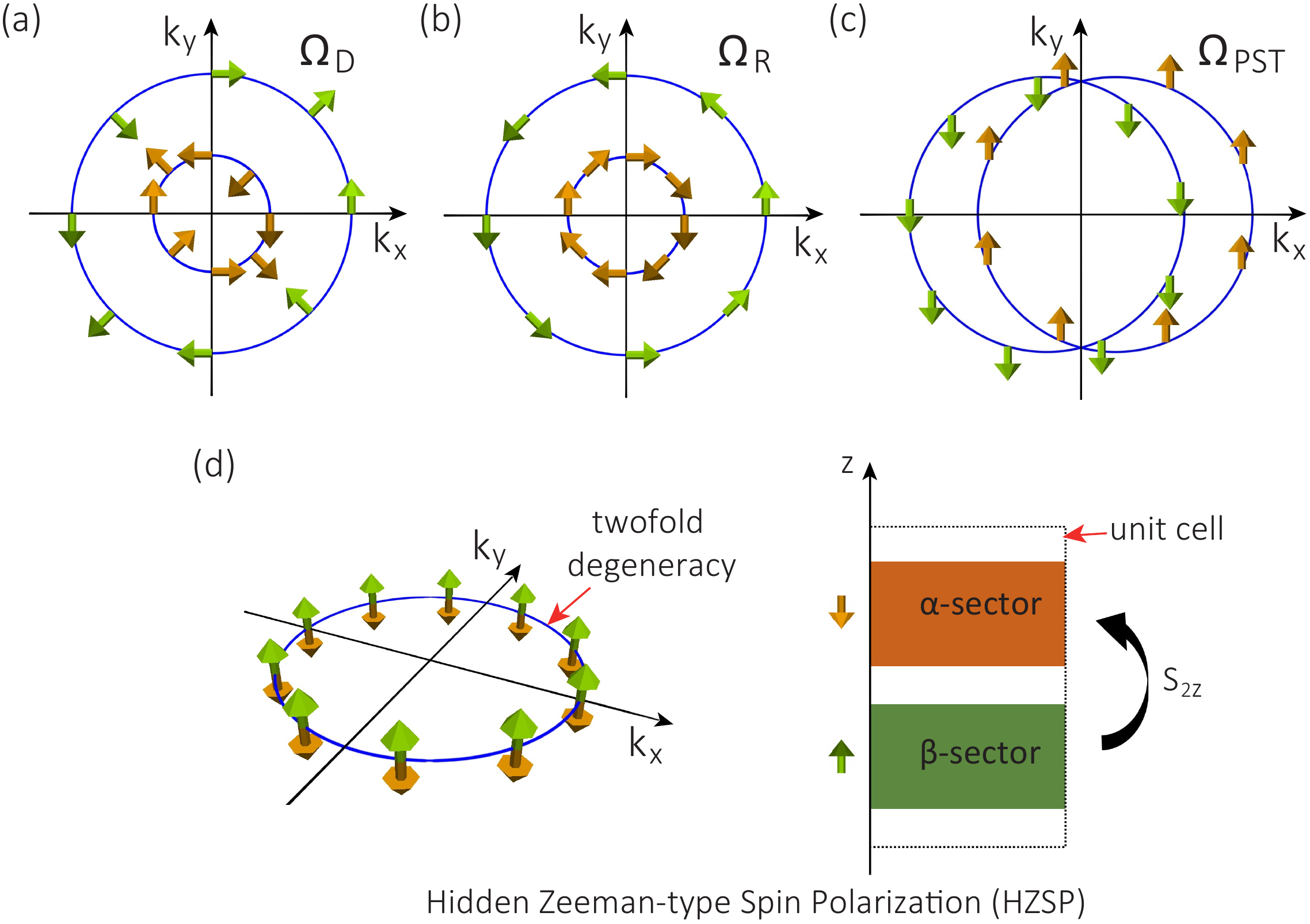}}
\caption{Spin texture in a QW system with (a) Dresselhaus, (b) Rashba and (c) PST configurations. (d) Schematic illustration of HZSP in momentum space (left panel) and real space (right panel). The black dashed box in the right panel indicates a unit cell comprising $\alpha$- and $\beta$-sectors. The two sectors, corresponding to the separated real-space distribution of the two degenerate subband states, are connected by a $S_{2z}$ operation.} 
\label{fig1}
\end{figure}

The spin-orbit interaction plays a key role in many of the proposed hidden spin effects occurring in centrosymmetric crystals containing inversion-paired asymmetric sectors (or sublattices) since it entangles the spin and orbital degrees of freedom. The spin-orbit interaction in combination with asymmetric crystal field causes spin splitting in solids lacking an inversion center~\cite{Dresselhaus,Rashba1984}, thereby producing an effective momentum-dependent magnetic field $\bm \Omega(\bm k)$ that couples to spin $\bm \sigma$~\cite{Manchon2015,Winkler2003}. This effect is known as Dresselhaus spin-orbit coupling (SOC), if the asymmetric crystal field arises from bulk-inversion asymmetry, or Rashba SOC, if the asymmetric crystal field is ascribed to structural inversion asymmetry in heterostructures or intrinsic electric dipoles in bulks. Therefore, the crystal space symmetry dictates the specific form of the effective magnetic field as well as the pattern of the corresponding spin texture. Taking a [001]-oriented III-V zinc-blende semiconductor quantum well (QW) with the $C_{2v}$ point group as an example~\cite{Winkler2003,Tao2017}, the effective magnetic field of the Dresselhaus effect $\bm \Omega_D(\bm k)$ takes the form of $\lambda_D(k_y,k_x)$ [see Fig.~\ref{fig1}(a)], whereas that of Rashba effect $\bm \Omega_R(\bm k)$ is written as $\lambda_R(-k_y,k_x)$, corresponding to a characteristic helical spin texture [see Fig.~\ref{fig1}(b)]. These chiral spin textures driven by the SOC have been harnessed to create non-equilibrium spin polarization and explored for kinds of spintronics applications, such as spin-Hall conductivity~\cite{Hirsch1999,Zhang2000}, spin-galvanic effect~\cite{Ganichev2002}, current-induced spin polarization~\cite{Edelstein1990}, and spin field-effect transistors~\cite{Datta1990}. It led to the emergence of a branch of spintronics---spin-orbitronics~\cite{Manchon2015}.
 
Specially, a persistent spin texture (PST)~\cite{Schliemann2017a} emerges in e.g., III-V zinc-blende QWs if $\lambda_D=\pm \lambda_R=\lambda/2$, which yields a unidirectional effective magnetic field $\bm \Omega_{PST}$ described by $\lambda(k_y,0)$ or $\lambda(0,k_x)$ and guarantees to maintain a uniform spin configuration in momentum space~\cite{Tao2018}. Figure~\ref{fig1}(c) schematically shows that each subband possesses a uniform spin configuration with spin orientation exactly pointing to the $+y$- or $-y$-direction as a result of $\bm \Omega_{PST}=\lambda(0,k_x)$. It envisions a possibility to realize a spatially periodic mode, i.e., persistent spin helix (PSH)~\cite{Bernevig2006}. The PSH state is protected by SU(2) spin rotation symmetry and thus is robust against spin-independent disorder scattering, which is crucial for achieving a long spin lifetime~\cite{Kammermeier2016}. However, it has only been experimentally confirmed in very few semiconductor QWs~\cite{Koralek2009,Walser2012,Kohda2012,Sasaki2014}.

Inspired by the discovery of hidden Rashba and Dresselhaus effects, one may wonder whether the PST could also exist but be hidden in solid-state materials. More interestingly, the hidden persistent spin texture (HPST) should be an intrinsic property that circumvents the stringent requirements for achieving $\lambda_D=\pm \lambda_R$ in semiconductor QW systems (e.g., the precise control of the QW width, the strength of the external electric field, and the doping level~\cite{Schliemann2017a}), going beyond the existing design paradigm of PST. 

In this work, we uncover the existence of HPST with an out-of-plane spin configuration, thus dubbed as hidden Zeeman-type spin polarization (HZSP), in bulk crystals containing sublattices by using a symmetry analysis in conjunction with density-functional theory (DFT) calculations. We find that compared with conventional PST for spin-splitting states, HZSP exhibits a unique spin-sector locking effect, as schematically shown in Fig.~\ref{fig1}(d): the momentum-independent uniform spin orientations of one spin-subband are compensated completely by the opposite spin orientation of the remaining spin-subband for a spin-degenerate band; whereas, in real space, opposite spins are spatially segregated in two paired sectors. We further reveal that such a HZSP is enforced by a nonsymmorphic twofold screw-rotational operation, i.e., $S_{2z}:(x,y,z)\rightarrow(-x,-y,z+\frac{1}{2})$ (here the screw axis is taken along $z$ for simplicity), which acts on the sublattice degree of freedom by connecting the $\alpha$- to $\beta$-sector.

\emph{{\color{blue} HZSP in centrosymmetric crystals.}} We first consider the centrosymmetric systems. Here, both the space inversion symmetry $\mathcal{P}$ and time-reversal symmetry $\mathcal{T}$ are preserved, and thereby we have $(\mathcal{PT})^2=-1$ for a spin-half system, ensuring all bands to be, at least, twofold degenerate in the whole Brillouin zone (BZ). The degenerate doublet states at each $\bm k$ point form a Kramers pair ($\psi_k,\Theta\psi_k$), where $\Theta=\mathcal{PT}$. However, this joint operation does not impose enough constraint on the specific form of $\bm \Omega(\bm k)$ to produce an out-of-plane PST for each local sector, which instead can be enforced additionally by the nonsymmorphic symmetry $S_{2z}$. To be specific, $S_{2z}$ combined with $\mathcal{P}$ generates a glide mirror operation, which can produce the PST in momentum space~\cite{Ji2022}. More importantly, $S_{2z}$ further guarantees the states of energy bands on the BZ surface ($k_z=\pi$) segregated on one of inversion-paired sublattices, giving rise to the symmetry-protected hidden spin polarization~\cite{Yuan2019,Guan2022}. Therefore, centrosymmetric crystals contain $S_{2z}$ operations are supposed to exhibit HZSP.

{\it Detailed symmetry analysis.} We focus on the $k_z=\pi$ plane, and \emph{any} point on this plane is invariant under the combined operation $S_{2z}\mathcal{P}$, i.e., $\widetilde{M}_z:(x,y,z)\rightarrow(x,y,-z+\frac{1}{2})$. Because $(S_{2z}\mathcal{P})^2=\widetilde{M}_z^2=-1$ at $k_z=\pi$ considering also the spin space, we can label the doublet ($\psi_k,\Theta\psi_k$) by using the eigenvalues $\pm i$ of $\widetilde{M}_z$, namely, $\widetilde{M}_z|\psi_k^{\pm i} \rangle=\pm i|\psi_k^{\pm i} \rangle$ and $\widetilde{M}_z|\Theta\psi_k^{\pm i} \rangle=\pm i|\Theta\psi_k^{\pm i} \rangle$. As a result, two conjugated doublets ($\psi_k^{+i},\Theta\psi_k^{+i}$) and ($\psi_k^{-i},\Theta\psi_k^{-i}$) with different $\widetilde{M}_z$ eigenvalues appear at each point on that plane. 

Since $\widetilde{M}_z$ anticommutes with $\sigma_x$ and $\sigma_y$ in the spin space ($\bm s=\frac{\hbar}{2}\langle\bm\sigma\rangle$, where ${\bm \sigma}=(\sigma_x,\sigma_y,\sigma_z)$ and $\sigma_x, \sigma_y, \sigma_z$ are Pauli matrices), we have $\langle \psi_k^{+ i}|\sigma_{x,y}|\psi_k^{+ i} \rangle=\langle \psi_k^{+ i}|(\widetilde{M}_z)^{-1}\sigma_{x,y}\widetilde{M}_z|\psi_k^{+ i} \rangle=-\langle \psi_k^{+ i}|\sigma_{x,y}|\psi_k^{+ i} \rangle$, which means $\langle \psi_k^{+ i}|\sigma_{x,y}|\psi_k^{+ i} \rangle=0$. A similar analysis results in $\langle \Theta\psi_k^{+ i}|\sigma_{x,y}|\Theta\psi_k^{+ i} \rangle=0$. The same conclusion also applies to the other doublet ($\psi_k^{-i},\Theta\psi_k^{-i}$). That is to say, the expectation values of spin operators $s_x$ and $s_y$ are forced to be zero within each of the two doublets. In contrast, the $z$-component spin expectations ($\langle s_{z}\rangle$) are finite but with opposite orientations between two states of each degenerated doublet as ensured by $\Theta$. In addition,  the combination of $S_{2z}$ and $\mathcal{T}$ acting on the real space enforces the wavefunction segregation of such doublet states in one of the paired sectors linked by $S_{2z}$~\cite{Guan2022}. Therefore, a symmetry-enforced uniform spin configuration pointing to the $z$-direction locally in each sector is realized on the $k_z=\pi$ plane, manifesting as a HZSP.

\begin{figure}[!ht]
\centerline{\includegraphics[width=0.5\textwidth]{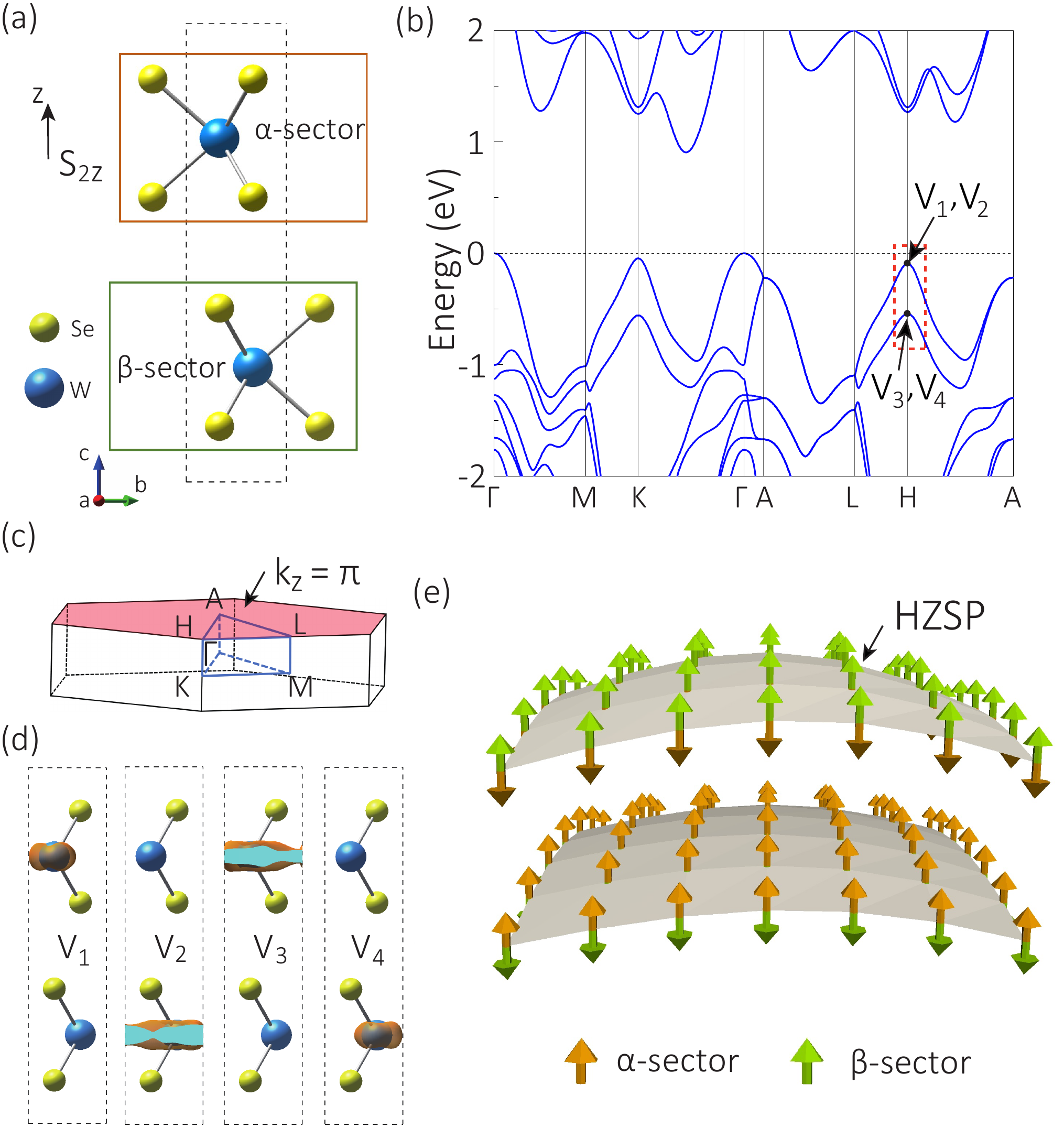}}
\caption{(a) Crystal structure of centrosymmetric bulk WSe$_2$, which contains a twofold screw rotation $S_{2z}$. (b) Band structure in the presence of SOC. (c) The corresponding BZ, where the $k_z=\pi$ plane is schematically shown by the red shaded surfaces. (d) Charge density distribution plotted for states V$_1$, V$_2$, V$_3$ and V$_4$. (e) HZSP around H indicated by the red box in (b).}
\label{fig2}
\end{figure}

{\it Possible candidates.} To explicitly demonstrate the above point in the case of centrosymmetric systems, we carry out a DFT calculation on a actual material WSe$_2$~\cite{sp}. As illustrated in Fig.~\ref{fig2}(a), it has a van der Waals layered crystal structure in the space group $P6_{3}/mmc$ (No. 194)~\cite{Coehoorn1987}, which contains an inversion center and a screw axis $S_{2z}$. In one unit cell, two sandwich layers, which are related to each other by $S_{2z}$, are indicated as the $\alpha$-sector and $\beta$-sector. Fig.~\ref{fig2}(b) depicts the calculated band structure along the high symmetry lines in the BZ with SOC included. The material has an indirect bandgap of 0.91 eV with the conduction-band minimum (CBM) located on the $K-\Gamma$ path, whereas the valence-band maximum (VBM) is at the $\Gamma$ point. Besides, each band is doubly degenerate and no spin splitting is observed. 

According to the above analysis, there must exists a HZSP located at the $k_z=\pi$ plane containing the route $A-L-H-A$, as shown by the red shaded surface in Fig.~\ref{fig2}(c). Our first-principles calculation results indeed confirm it~\cite{sp}. As a typical example, we plot the two-dimensional (2D) band dispersion together with the calculated local spin polarizations in the vicinity of the H point when $k_z=\pi$ [see Fig.~\ref{fig2}(e)]. It clearly shows that each of the two twofold degenerate band states have finite and opposite spin polarizations, whose orientations are exactly along the $z$-direction and remain independent of the momentum. Meanwhile, by examining the spatial distribution of the two valence-band doublet states marked by (V$_1$, V$_2$) and (V$_3$, V$_4$) at the H point [see Fig.~\ref{fig2}(d)], one finds the following features. (i) The band states are mainly distributed in the two Mo atomic layers. (ii) The charge density distribution of the V$_1$ state is confined in the $\alpha$-sector whereas for the V$_2$ state it is segregated in the $\beta$-sector. (iii) The sector distribution pattern is reversed for the other doublet states V$_3$ and V$_4$. Therefore, the first-principles calculation identifies the existence of HZSP in WSe$_2$, consistent with our symmetry analysis.

\begin{figure*}[!htb]
\centerline{\includegraphics[width=0.88\textwidth]{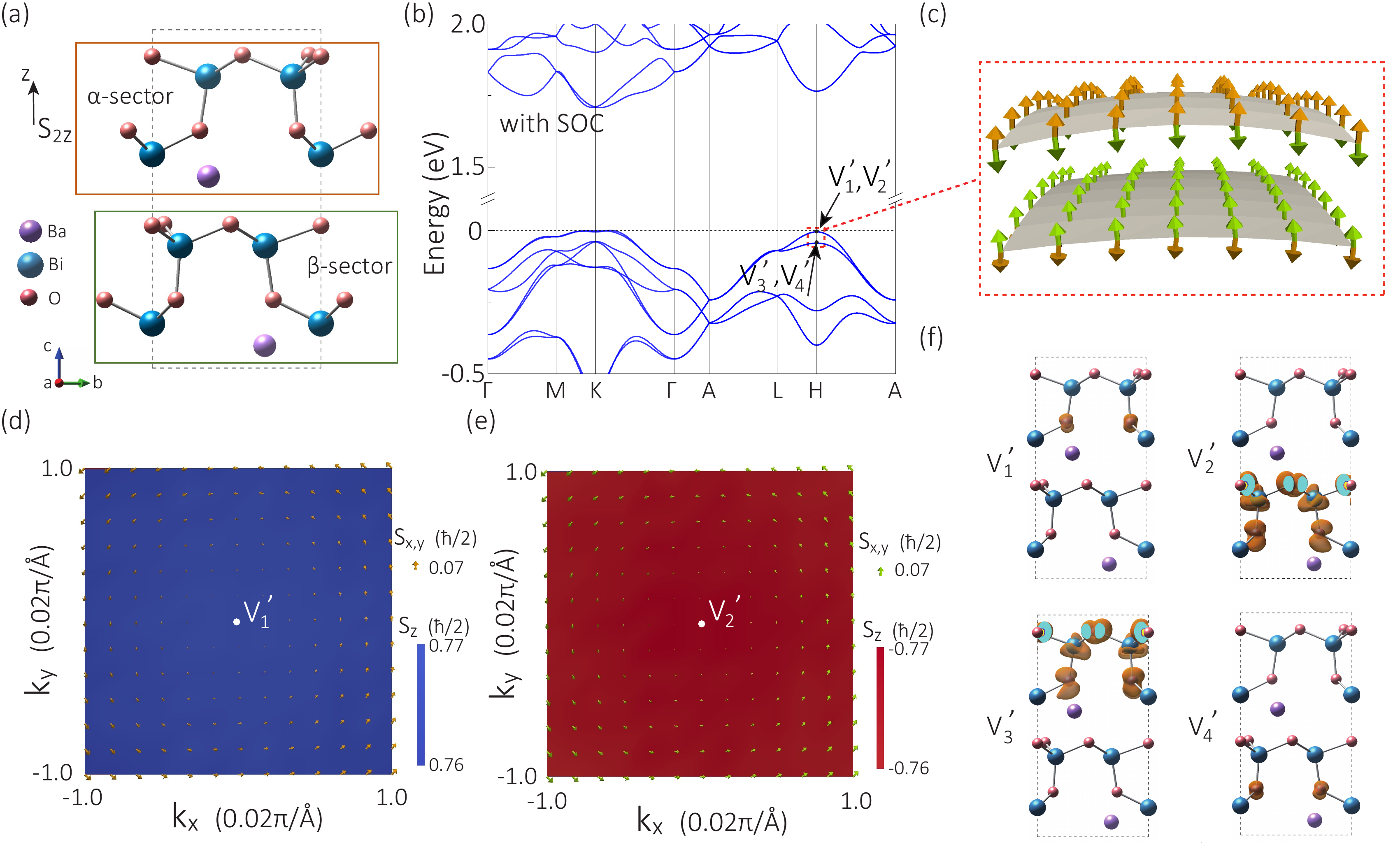}}
\caption{(a) Crystal structure of non-centrosymmetric BaBi$_4$O$_7$, where the nonsymmorphic $S_{2z}$ is preserved. (b) Band structure for BaBi$_4$O$_7$ with SOC included. (c) HZSP around the H point indicated by the red dashed box in (b). (d-e) 2D plot of spin patterns for the two highest occupied doublet states around H. The in-plane spin components $s_x$ and $s_y$, are represented by the arrows whereas the out-of-plane spin component $s_z$ is shown by color. (f) Charge density distribution plotted for states V$_1^\prime$, V$_2^\prime$, V$_3^\prime$ and V$_4^\prime$. }
\label{fig3}
\end{figure*}

 It is worth pointing out that the proposed HZSP here shares similar features as the D-2 effect of NaCaBi in Ref.~\cite{Zhang2014}, where the spin pattern in the vicinity of the band crossing points is simply shown with a symmetry analysis only in terms of the site point group performed. The above discussion indicates that HZSP is able to cover the whole plane of the BZ boundary. Besides, the underlying mechanism is attributed to the nonsymmorphic operation $S_{2z}$, which plays a more important role than the symmorphic one. Interestingly, thanks to the protection of $S_{2z}$, HZSP can induce intrinsic circular polarization as high as 100\% at the H valley in WSe$_2$, which is larger than that (70\%) at the K valley~\cite{Liu2015}. In particular, the circular polarization at H is insensitive to interlayer distance, demonstrating a potential optoelectronic application based on HZSP.

\emph{{\color{blue} HZSP in noncentrosymmetric crystals.}} Is it possible to realize HZSP in inversion-asymmetric materials? Now, the space inversion symmetry $\mathcal{P}$ is absent, so that we have to find a new way to guarantee the symmetry-protected double degeneracy in comparison with centrosymmetric cases. We find the combination of a $S_{2z}$ with the retained $\mathcal{T}$ can meet this objective despite the broken $\mathcal{P}$. The reason is that the $S_{2z}\mathcal{T}$ operation preserves the $\bm k$ point on the $k_z=\pi$ plane and $(S_{2z}\mathcal{T})^2=-1$, which can generate Kramers doublet states ($\psi_k,\widetilde{\Theta}\psi_k$), where $\widetilde{\Theta}=S_{2z}\mathcal{T}$, and thereby can protect the states on a whole BZ surface with a twofold degeneracy, known as nodal surface states~\cite{Zhong2016,Liang2016,Wu2018}. On this basis, we propose that the minimal set of symmetries should include another n-fold rotational operation ($n\geqslant3$) along the $z$-direction to enable a HZSP appearing around certain high symmetry points (not limited to time-reversal invariant momentum point).

{\it Possible candidates.} We then illustrate this behavior via a concrete material example--- BaBi$_4$O$_7$. It crystallizes in a hexagonal structure with space group P6$_3$mc (No. 186)\cite{Klinkova1996}, which excludes the inversion center but maintains $S_{2z}$. Figure~\ref{fig3}(a) shows the crystal structure of BaBi$_4$O$_7$, of which the unit cell can also be divided into a pair of sectors (termed as $\alpha$-sector and $\beta$-sector) connected by $S_{2z}$. The calculated band structure for BaBi$_4$O$_7$ in the presence of SOC is plotted in Fig.~\ref{fig3}(b). One observes that it is an indirect bandgap semiconductor with a fundamental band gap of 1.71 eV. As expected, the absence of $\mathcal{P}$ generally lifts the band degeneracy in the BZ except for the band states on the $k_z=\pi$ plane, e.g., along the high symmetry lines $A-L-H-A$ (see the supplement material~\cite{sp} for the band structure without considering SOC). This is in agreement with the above argument that the antiunitary symmetry $\widetilde{\Theta}$ guarantees a twofold degeneracy on that plane.

Similar to the case of WSe$_2$, we focus our attention on $\bm k$ points around the high symmetry point H. The calculated 2D band structure and the corresponding sector-decomposed spin polarization are shown in Fig.~\ref{fig3}(c). Interestingly, we do, on the whole, obtain a HZSP-like pattern with a strongly dominant $z$-component of the spin vector in the BZ. Moreover, we plot the charge density distribution of the hole states indicated by V$_1^\prime$, V$_2^\prime$, V$_3^\prime$, V$_4^\prime$ around the H point in Fig.~\ref{fig3}(f), which share similar features to those of WSe$_2$, further confirming the emergent HZSP in BaBi$_4$O$_7$. It is noteworthy that such an unexpected hidden spin pattern in noncentrosymmetric materials has yet to be reported.

Figures~\ref{fig3}(d-e) illustrate the 2D diagrams of the spin polarizations for nodal surface doublet states ($\psi_k,\widetilde{\Theta}\psi_k$) near the H point, where the sampled centers correspond to the states V$_1^\prime$ and V$_2^\prime$, respectively. In comparison with the HZSP in centrosymmetric crystals, the main difference lies in that there is an increasing in-plane spin component as the momentum moves away from the H point, which cannot be compensated by its Kramers partner. In contrast, the doublet states have the same in-plane magnitude and orientation of the spin vector. This can be easily understood by the antiunitary symmetry $\widetilde{\Theta}$ acting on the spin space. In this process, the $\mathcal{T}$ first transforms the spin components from ($s_x$, $s_y$, $s_z$) to (-$s_x$, -$s_y$, -$s_z$), and then $S_{2z}$ transforms (-$s_x$, -$s_y$, -$s_z$) to ($s_x$, $s_y$, -$s_z$), leading to the opposite out-of-plane spin components but the same in-plane spin components between $\psi_k$ and $\widetilde{\Theta}\psi_k$. Additionally, we see that the in-plane spin components $s_{x,y}$ exhibit a helical spin texture with a counterclockwise chirality.

Next, we construct an effective $k\cdot p$ Hamiltonian to further explain the unique spin textures occurring in BaBi$_4$O$_7$. The symmetry operations in the little group at H include $S_{2z}\mathcal{T}$, $C_{3z}$ and additionally $\widetilde{M}_x:(x,y,z)\rightarrow(-x,y,z+\frac{1}{2})$, with the following matrix representations~\cite{Bradley1972}, $S_{2z}\mathcal{T}=i\tau_y \otimes\sigma_x \mathcal{K}$, $C_{3z}=\tau_0 \otimes e^{-i\frac{\pi}{3}\sigma_z}$ and $\widetilde{M}_x=\tau_y \otimes\sigma_x$. Here, $\tau$ and $\sigma$ are Pauli matrices describing the sublattice and spin spaces, respectively, and $\mathcal{K}$ is the complex conjugation. Constrained by these symmetries, the effective Hamiltonian to the lowest order expanded around H is given by 
\begin{equation}
\mathcal{H}(\bm k)=\alpha(k_y\sigma_x-k_x \sigma_y)+M\tau_z\otimes\sigma_z-\beta k_z\tau_y\otimes\sigma_0,
\label{equ1}
\end{equation}
where $M$, $\alpha$, and $\beta$ are constants, and the wave vector $\bm k$ is referenced with respect to the H point. The band energies of Eq.~\ref{equ1} read $E(\bm k)=\pm \sqrt{M^2+(\alpha k_{\parallel}\pm\beta k_z)^2}$, where $k_\parallel=\sqrt{k_x^2+k_y^2}$. 

For $k_z=0$, the Hamiltonian~(\ref{equ1}) is reduced to $\mathcal{H}(k_{\parallel})=\alpha(k_y\sigma_x-k_x \sigma_y)+M\tau_z\otimes\sigma_z$. One finds that there emerge two doublets with degenerate energies $E^{\pm}(k_{\parallel})=\pm \sqrt{M^2+(\alpha k_{\parallel})^2}$, yielding nodal surface states on the BZ boundary. Furthermore, the Hamiltonian now can be easily diagonalized in the sublattice space due to its commutativity with the sector measurement operator $\tau_z \otimes\sigma_0$. We then obtain an equivalent effective magnetic field $\bm \Omega(\bm k)$ taking the form of $(\alpha k_y,-\alpha k_x,\eta M)$, which has the opposite sign of the $z$-component for different sectors, i.e., the values of $\eta= \pm 1$. Accordingly, the expectation values of the spin operator for the two doublet states are calculated to be
\begin{equation}
(s_x, s_y, s_z)^{\pm}_{\eta}=\pm \frac{\hbar}{2\sqrt{M^2+(\alpha k_{\parallel})^2}}(-\alpha k_y,\alpha k_x,\eta M).
\label{equ2}
\end{equation}
Hence, there exists a totally compensated $z$-component of the spin vector for each doublet exactly at H (when $k_{\parallel}=0$). Besides, the in-plane components within each doublet share the same linear dependence on $k_{\parallel}$, suggesting that the momentum-space averaged spin moment is non-zero for the in-plane direction when $k_{\parallel}\ne0$. The results are in line with the DFT calculations. The evaluated parameters by fitting the calculation results are $M$ = 0.02 eV and $\alpha$ = 0.031 eV\AA. As a result, the in-plane spin components in the vicinity of the H point are negligible compared with its $z$-component, exhibiting a nearly HZSP that covers a large portion of the BZ. The range of $k_x$ and $k_y$ values in Figs.~\ref{fig3}(d-e) spans 0.04$\pi$ \AA$^{-1}$ around H, comparable to the values proposed in BiInO$_3$ with conventional PST ($\sim0.2$ $\AA^{-1}$)~\cite{Tao2018}. This makes BaBi$_4$O$_7$ ideal for studying HZSP in noncentrosymmetric materials.

\emph{{\color{blue} Discussion.}} We have uncovered the HZSP in both centrosymmetric and non-centrosymmetric bulk materials, where the nonsymmorphic symmetry plays a crucial role. As discussed above, the screw axis has been assumed along $z$, thereby displaying a HZSP at $k_z=\pi$. Similarly, if we have a screw axis along the direction $x$ or $y$, a HZSP at $k_x=\pi$ or $k_y=\pi$ can be realized correspondingly. Notably, the presence of HZSP discussed in centrosymmetric crystals is solely determined by symmetry, and its location is fixed at the BZ boundary plane. For this reason, the essentially symmetry-enforced HZSP is independent of the specific material, making it easier for the exploration of candidate materials by analyzing the space groups with an inversion center~\cite{Aroyo2006}. 

We propose that the HZSP behavior revealed in WSe$_2$ and BaBi$_4$O$_7$ can be directly probed via nuclear magnetic resonance (NMR) measurements~\cite{Ramirez-Ruiz2017} or spin- and angle-resolved photoemission spectroscopy~\cite{Riley2014,Wu2017a,Yao2017}. The former requires a uniform electric field to create a staggered magnetic field pointing to opposite directions at atomic sites of the  $S_{2z}$-connected sectors, resulting in the signals from the connected nuclei as two splitting NMR resonance peaks~\cite{Ramirez-Ruiz2017}. The latter works because a probing beam penetrating the sample is attenuated with depth, so that the average of spin polarization over depth is non-zero and thus is observable. Notably, during the photoemission process, the photon energy should be varied to be able to scan the out-of-plane momentum of the photoelectron, thereby determining the momentum exactly located at the BZ boundary~\cite{Gehlmann2016}. Particularly, our proposed HZSP here is gauge-invariant since the choice of sectors is definite thanks to the nontrival protection of wavefunction segregation by $S_{2z}$~\cite{Li2018,Yuan2019}.

We may expect interesting properties for states around HZSP such as the large anisotropic g-factor arising from the characteristic pattern of the effective magnetic field locked to a unique direction in centrosymmetric crystals. In addition, for HZSP in non-centrosymmetric materials, an enhanced mobility along the $z$-direction is also expected because of the suppressed backscattering from scatterers that preserve the spin.

\begin{acknowledgments}
This work is supported by the National Key Research and Development Program of China (Grant No. 2018YFB2202800), and the Science Foundation of CAS (Grant No. 2022000045). J.W.L. is supported by the National Natural Science Foundation of China (Grants No. 61888102 and No. 11925407), and the National Key Research and Development Program of China (Grant No. 2018YFB2200105). 
\end{acknowledgments}


\begin{thebibliography}{45}%
\makeatletter
\providecommand \@ifxundefined [1]{%
 \@ifx{#1\undefined}
}%
\providecommand \@ifnum [1]{%
 \ifnum #1\expandafter \@firstoftwo
 \else \expandafter \@secondoftwo
 \fi
}%
\providecommand \@ifx [1]{%
 \ifx #1\expandafter \@firstoftwo
 \else \expandafter \@secondoftwo
 \fi
}%
\providecommand \natexlab [1]{#1}%
\providecommand \enquote  [1]{``#1''}%
\providecommand \bibnamefont  [1]{#1}%
\providecommand \bibfnamefont [1]{#1}%
\providecommand \citenamefont [1]{#1}%
\providecommand \href@noop [0]{\@secondoftwo}%
\providecommand \href [0]{\begingroup \@sanitize@url \@href}%
\providecommand \@href[1]{\@@startlink{#1}\@@href}%
\providecommand \@@href[1]{\endgroup#1\@@endlink}%
\providecommand \@sanitize@url [0]{\catcode `\\12\catcode `\$12\catcode
  `\&12\catcode `\#12\catcode `\^12\catcode `\_12\catcode `\%12\relax}%
\providecommand \@@startlink[1]{}%
\providecommand \@@endlink[0]{}%
\providecommand \url  [0]{\begingroup\@sanitize@url \@url }%
\providecommand \@url [1]{\endgroup\@href {#1}{\urlprefix }}%
\providecommand \urlprefix  [0]{URL }%
\providecommand \Eprint [0]{\href }%
\providecommand \doibase [0]{http://dx.doi.org/}%
\providecommand \selectlanguage [0]{\@gobble}%
\providecommand \bibinfo  [0]{\@secondoftwo}%
\providecommand \bibfield  [0]{\@secondoftwo}%
\providecommand \translation [1]{[#1]}%
\providecommand \BibitemOpen [0]{}%
\providecommand \bibitemStop [0]{}%
\providecommand \bibitemNoStop [0]{.\EOS\space}%
\providecommand \EOS [0]{\spacefactor3000\relax}%
\providecommand \BibitemShut  [1]{\csname bibitem#1\endcsname}%
\let\auto@bib@innerbib\@empty
\bibitem [{\citenamefont {Zhang}\ \emph {et~al.}(2014)\citenamefont {Zhang},
  \citenamefont {Liu}, \citenamefont {Luo}, \citenamefont {Freeman},\ and\
  \citenamefont {Zunger}}]{Zhang2014}%
  \BibitemOpen
  \bibfield  {author} {\bibinfo {author} {\bibfnamefont {X.}~\bibnamefont
  {Zhang}}, \bibinfo {author} {\bibfnamefont {Q.}~\bibnamefont {Liu}}, \bibinfo
  {author} {\bibfnamefont {J.-W.}\ \bibnamefont {Luo}}, \bibinfo {author}
  {\bibfnamefont {A.~J.}\ \bibnamefont {Freeman}}, \ and\ \bibinfo {author}
  {\bibfnamefont {A.}~\bibnamefont {Zunger}},\ } {\bibfield  {journal} {\bibinfo
  {journal} {Nature Physics}\ }\textbf {\bibinfo {volume} {10}},\ \bibinfo
  {pages} {387} (\bibinfo {year} {2014})}\BibitemShut {NoStop}%
\bibitem [{\citenamefont {Partoens}(2014)}]{Partoens2014}%
  \BibitemOpen
  \bibfield  {author} {\bibinfo {author} {\bibfnamefont {B.}~\bibnamefont
  {Partoens}},\ } {\bibfield
  {journal} {\bibinfo  {journal} {Nature Physics}\ }\textbf {\bibinfo {volume}
  {10}},\ \bibinfo {pages} {333} (\bibinfo {year} {2014})}\BibitemShut
  {NoStop}%
\bibitem [{\citenamefont {Riley}\ \emph {et~al.}(2014)\citenamefont {Riley},
  \citenamefont {Mazzola}, \citenamefont {Dendzik}, \citenamefont {Michiardi},
  \citenamefont {Takayama}, \citenamefont {Bawden}, \citenamefont {Granerd},
  \citenamefont {Leandersson}, \citenamefont {Balasubramanian}, \citenamefont
  {Hoesch}, \citenamefont {Kim}, \citenamefont {Takagi}, \citenamefont
  {Meevasana}, \citenamefont {Hofmann}, \citenamefont {Bahramy}, \citenamefont
  {Wells},\ and\ \citenamefont {King}}]{Riley2014}%
  \BibitemOpen
  \bibfield  {author} {\bibinfo {author} {\bibfnamefont {J.~M.}\ \bibnamefont
  {Riley}}, \bibinfo {author} {\bibfnamefont {F.}~\bibnamefont {Mazzola}},
  \bibinfo {author} {\bibfnamefont {M.}~\bibnamefont {Dendzik}}, \bibinfo
  {author} {\bibfnamefont {M.}~\bibnamefont {Michiardi}}, \bibinfo {author}
  {\bibfnamefont {T.}~\bibnamefont {Takayama}}, \bibinfo {author}
  {\bibfnamefont {L.}~\bibnamefont {Bawden}}, \bibinfo {author} {\bibfnamefont
  {C.}~\bibnamefont {Granerd}}, \bibinfo {author} {\bibfnamefont
  {M.}~\bibnamefont {Leandersson}}, \bibinfo {author} {\bibfnamefont
  {T.}~\bibnamefont {Balasubramanian}}, \bibinfo {author} {\bibfnamefont
  {M.}~\bibnamefont {Hoesch}}, \bibinfo {author} {\bibfnamefont {T.~K.}\
  \bibnamefont {Kim}}, \bibinfo {author} {\bibfnamefont {H.}~\bibnamefont
  {Takagi}}, \bibinfo {author} {\bibfnamefont {W.}~\bibnamefont {Meevasana}},
  \bibinfo {author} {\bibfnamefont {P.}~\bibnamefont {Hofmann}}, \bibinfo
  {author} {\bibfnamefont {M.}~\bibnamefont {Bahramy}}, \bibinfo {author}
  {\bibfnamefont {J.}~\bibnamefont {Wells}}, \ and\ \bibinfo {author}
  {\bibfnamefont {P.~C.}\ \bibnamefont {King}},\ } {\bibfield  {journal} {\bibinfo
  {journal} {Nature Physics}\ }\textbf {\bibinfo {volume} {10}},\ \bibinfo
  {pages} {835} (\bibinfo {year} {2014})}\BibitemShut {NoStop}%
\bibitem [{\citenamefont {Wu}\ \emph {et~al.}(2017)\citenamefont {Wu},
  \citenamefont {Sumida}, \citenamefont {Miyamoto}, \citenamefont {Taguchi},
  \citenamefont {Yoshikawa}, \citenamefont {Kimura}, \citenamefont {Ueda},
  \citenamefont {Arita}, \citenamefont {Nagao}, \citenamefont {Watauchi},
  \citenamefont {Tanaka},\ and\ \citenamefont {Okuda}}]{Wu2017a}%
  \BibitemOpen
  \bibfield  {author} {\bibinfo {author} {\bibfnamefont {S.-L.}\ \bibnamefont
  {Wu}}, \bibinfo {author} {\bibfnamefont {K.}~\bibnamefont {Sumida}}, \bibinfo
  {author} {\bibfnamefont {K.}~\bibnamefont {Miyamoto}}, \bibinfo {author}
  {\bibfnamefont {K.}~\bibnamefont {Taguchi}}, \bibinfo {author} {\bibfnamefont
  {T.}~\bibnamefont {Yoshikawa}}, \bibinfo {author} {\bibfnamefont
  {A.}~\bibnamefont {Kimura}}, \bibinfo {author} {\bibfnamefont
  {Y.}~\bibnamefont {Ueda}}, \bibinfo {author} {\bibfnamefont {M.}~\bibnamefont
  {Arita}}, \bibinfo {author} {\bibfnamefont {M.}~\bibnamefont {Nagao}},
  \bibinfo {author} {\bibfnamefont {S.}~\bibnamefont {Watauchi}}, \bibinfo
  {author} {\bibfnamefont {I.}~\bibnamefont {Tanaka}}, \ and\ \bibinfo {author}
  {\bibfnamefont {T.}~\bibnamefont {Okuda}},\ } {\bibfield  {journal} {\bibinfo
  {journal} {Nature Communications}\ }\textbf {\bibinfo {volume} {8}},\
  \bibinfo {pages} {1919} (\bibinfo {year} {2017})}\BibitemShut {NoStop}%
\bibitem [{\citenamefont {Yao}\ \emph {et~al.}(2017)\citenamefont {Yao},
  \citenamefont {Wang}, \citenamefont {Huang}, \citenamefont {Deng},
  \citenamefont {Yan}, \citenamefont {Zhang}, \citenamefont {Miyamoto},
  \citenamefont {Okuda}, \citenamefont {Li}, \citenamefont {Wang},
  \citenamefont {Gao}, \citenamefont {Liu}, \citenamefont {Duan},\ and\
  \citenamefont {Zhou}}]{Yao2017}%
  \BibitemOpen
  \bibfield  {author} {\bibinfo {author} {\bibfnamefont {W.}~\bibnamefont
  {Yao}}, \bibinfo {author} {\bibfnamefont {E.}~\bibnamefont {Wang}}, \bibinfo
  {author} {\bibfnamefont {H.}~\bibnamefont {Huang}}, \bibinfo {author}
  {\bibfnamefont {K.}~\bibnamefont {Deng}}, \bibinfo {author} {\bibfnamefont
  {M.}~\bibnamefont {Yan}}, \bibinfo {author} {\bibfnamefont {K.}~\bibnamefont
  {Zhang}}, \bibinfo {author} {\bibfnamefont {K.}~\bibnamefont {Miyamoto}},
  \bibinfo {author} {\bibfnamefont {T.}~\bibnamefont {Okuda}}, \bibinfo
  {author} {\bibfnamefont {L.}~\bibnamefont {Li}}, \bibinfo {author}
  {\bibfnamefont {Y.}~\bibnamefont {Wang}}, \bibinfo {author} {\bibfnamefont
  {H.}~\bibnamefont {Gao}}, \bibinfo {author} {\bibfnamefont {C.}~\bibnamefont
  {Liu}}, \bibinfo {author} {\bibfnamefont {W.}~\bibnamefont {Duan}}, \ and\
  \bibinfo {author} {\bibfnamefont {S.}~\bibnamefont {Zhou}},\ } {\bibfield  {journal} {\bibinfo
  {journal} {Nature Communications}\ }\textbf {\bibinfo {volume} {8}},\
  \bibinfo {pages} {14216} (\bibinfo {year} {2017})}\BibitemShut {NoStop}%
\bibitem [{\citenamefont {Yuan}\ \emph {et~al.}(2019)\citenamefont {Yuan},
  \citenamefont {Liu}, \citenamefont {Zhang}, \citenamefont {Luo},
  \citenamefont {Li},\ and\ \citenamefont {Zunger}}]{Yuan2019}%
  \BibitemOpen
  \bibfield  {author} {\bibinfo {author} {\bibfnamefont {L.}~\bibnamefont
  {Yuan}}, \bibinfo {author} {\bibfnamefont {Q.}~\bibnamefont {Liu}}, \bibinfo
  {author} {\bibfnamefont {X.}~\bibnamefont {Zhang}}, \bibinfo {author}
  {\bibfnamefont {J.-W.}\ \bibnamefont {Luo}}, \bibinfo {author} {\bibfnamefont
  {S.-S.}\ \bibnamefont {Li}}, \ and\ \bibinfo {author} {\bibfnamefont
  {A.}~\bibnamefont {Zunger}},\ } {\bibfield  {journal} {\bibinfo
  {journal} {Nature Communications}\ }\textbf {\bibinfo {volume} {10}},\
  \bibinfo {pages} {906} (\bibinfo {year} {2019})}\BibitemShut {NoStop}%
\bibitem [{\citenamefont {Lin}\ \emph {et~al.}(2020)\citenamefont {Lin},
  \citenamefont {Wang}, \citenamefont {Xu},\ and\ \citenamefont
  {Duan}}]{Lin2020}%
  \BibitemOpen
  \bibfield  {author} {\bibinfo {author} {\bibfnamefont {Z.}~\bibnamefont
  {Lin}}, \bibinfo {author} {\bibfnamefont {C.}~\bibnamefont {Wang}}, \bibinfo
  {author} {\bibfnamefont {Y.}~\bibnamefont {Xu}}, \ and\ \bibinfo {author}
  {\bibfnamefont {W.}~\bibnamefont {Duan}},\ } {\bibfield  {journal} {\bibinfo  {journal}
  {Phys. Rev. B}\ }\textbf {\bibinfo {volume} {102}},\ \bibinfo {pages}
  {165143} (\bibinfo {year} {2020})}\BibitemShut {NoStop}%
\bibitem [{\citenamefont {Liu}\ \emph {et~al.}(2015)\citenamefont {Liu},
  \citenamefont {Zhang},\ and\ \citenamefont {Zunger}}]{Liu2015}%
  \BibitemOpen
  \bibfield  {author} {\bibinfo {author} {\bibfnamefont {Q.}~\bibnamefont
  {Liu}}, \bibinfo {author} {\bibfnamefont {X.}~\bibnamefont {Zhang}}, \ and\
  \bibinfo {author} {\bibfnamefont {A.}~\bibnamefont {Zunger}},\ } {\bibfield  {journal} {\bibinfo
  {journal} {Phys. Rev. Lett.}\ }\textbf {\bibinfo {volume} {114}},\ \bibinfo
  {pages} {087402} (\bibinfo {year} {2015})}\BibitemShut {NoStop}%
\bibitem [{\citenamefont {Ryoo}\ and\ \citenamefont {Park}(2017)}]{Ryoo2017}%
  \BibitemOpen
  \bibfield  {author} {\bibinfo {author} {\bibfnamefont {J.~H.}\ \bibnamefont
  {Ryoo}}\ and\ \bibinfo {author} {\bibfnamefont {C.-H.}\ \bibnamefont
  {Park}},\ } {\bibfield  {journal}
  {\bibinfo  {journal} {NPG Asia Materials}\ }\textbf {\bibinfo {volume} {9}},\
  \bibinfo {pages} {e382} (\bibinfo {year} {2017})}\BibitemShut {NoStop}%
\bibitem [{\citenamefont {Cho}\ \emph {et~al.}(2018)\citenamefont {Cho},
  \citenamefont {Park}, \citenamefont {Hong}, \citenamefont {Jung},
  \citenamefont {Kim}, \citenamefont {Han}, \citenamefont {Kyung},
  \citenamefont {Kim}, \citenamefont {Mo}, \citenamefont {Denlinger},
  \citenamefont {Shim}, \citenamefont {Han}, \citenamefont {Kim},\ and\
  \citenamefont {Park}}]{Cho2018}%
  \BibitemOpen
  \bibfield  {author} {\bibinfo {author} {\bibfnamefont {S.}~\bibnamefont
  {Cho}}, \bibinfo {author} {\bibfnamefont {J.-H.}\ \bibnamefont {Park}},
  \bibinfo {author} {\bibfnamefont {J.}~\bibnamefont {Hong}}, \bibinfo {author}
  {\bibfnamefont {J.}~\bibnamefont {Jung}}, \bibinfo {author} {\bibfnamefont
  {B.~S.}\ \bibnamefont {Kim}}, \bibinfo {author} {\bibfnamefont
  {G.}~\bibnamefont {Han}}, \bibinfo {author} {\bibfnamefont {W.}~\bibnamefont
  {Kyung}}, \bibinfo {author} {\bibfnamefont {Y.}~\bibnamefont {Kim}}, \bibinfo
  {author} {\bibfnamefont {S.-K.}\ \bibnamefont {Mo}}, \bibinfo {author}
  {\bibfnamefont {J.~D.}\ \bibnamefont {Denlinger}}, \bibinfo {author}
  {\bibfnamefont {J.~H.}\ \bibnamefont {Shim}}, \bibinfo {author}
  {\bibfnamefont {J.~H.}\ \bibnamefont {Han}}, \bibinfo {author} {\bibfnamefont
  {C.}~\bibnamefont {Kim}}, \ and\ \bibinfo {author} {\bibfnamefont {S.~R.}\
  \bibnamefont {Park}},\ }
  {\bibfield  {journal} {\bibinfo  {journal} {Phys. Rev. Lett.}\ }\textbf
  {\bibinfo {volume} {121}},\ \bibinfo {pages} {186401} (\bibinfo {year}
  {2018})}\BibitemShut {NoStop}%
\bibitem [{\citenamefont {Liu}(2017)}]{Liu2017}%
  \BibitemOpen
  \bibfield  {author} {\bibinfo {author} {\bibfnamefont {C.-X.}\ \bibnamefont
  {Liu}},\ } {\bibfield
  {journal} {\bibinfo  {journal} {Phys. Rev. Lett.}\ }\textbf {\bibinfo
  {volume} {118}},\ \bibinfo {pages} {087001} (\bibinfo {year}
  {2017})}\BibitemShut {NoStop}%
\bibitem [{\citenamefont {Gotlieb}\ \emph {et~al.}(2018)\citenamefont
  {Gotlieb}, \citenamefont {Lin}, \citenamefont {Serbyn}, \citenamefont
  {Zhang}, \citenamefont {Smallwood}, \citenamefont {Jozwiak}, \citenamefont
  {Eisaki}, \citenamefont {Hussain}, \citenamefont {Vishwanath},\ and\
  \citenamefont {Lanzara}}]{Gotlieb2018}%
  \BibitemOpen
  \bibfield  {author} {\bibinfo {author} {\bibfnamefont {K.}~\bibnamefont
  {Gotlieb}}, \bibinfo {author} {\bibfnamefont {C.-Y.}\ \bibnamefont {Lin}},
  \bibinfo {author} {\bibfnamefont {M.}~\bibnamefont {Serbyn}}, \bibinfo
  {author} {\bibfnamefont {W.}~\bibnamefont {Zhang}}, \bibinfo {author}
  {\bibfnamefont {C.~L.}\ \bibnamefont {Smallwood}}, \bibinfo {author}
  {\bibfnamefont {C.}~\bibnamefont {Jozwiak}}, \bibinfo {author} {\bibfnamefont
  {H.}~\bibnamefont {Eisaki}}, \bibinfo {author} {\bibfnamefont
  {Z.}~\bibnamefont {Hussain}}, \bibinfo {author} {\bibfnamefont
  {A.}~\bibnamefont {Vishwanath}}, \ and\ \bibinfo {author} {\bibfnamefont
  {A.}~\bibnamefont {Lanzara}},\ }
  {\bibfield  {journal} {\bibinfo  {journal} {Science}\ }\textbf {\bibinfo
  {volume} {362}},\ \bibinfo {pages} {1271} (\bibinfo {year} {2018})}, \BibitemShut
  {NoStop}%
\bibitem [{\citenamefont {Yu}\ \emph {et~al.}(2020)\citenamefont {Yu},
  \citenamefont {Guan}, \citenamefont {Sheng}, \citenamefont {Gao},\ and\
  \citenamefont {Yang}}]{Yu2020}%
  \BibitemOpen
  \bibfield  {author} {\bibinfo {author} {\bibfnamefont {Z.-M.}\ \bibnamefont
  {Yu}}, \bibinfo {author} {\bibfnamefont {S.}~\bibnamefont {Guan}}, \bibinfo
  {author} {\bibfnamefont {X.-L.}\ \bibnamefont {Sheng}}, \bibinfo {author}
  {\bibfnamefont {W.}~\bibnamefont {Gao}}, \ and\ \bibinfo {author}
  {\bibfnamefont {S.~A.}\ \bibnamefont {Yang}},\ } {\bibfield  {journal} {\bibinfo  {journal}
  {Phys. Rev. Lett.}\ }\textbf {\bibinfo {volume} {124}},\ \bibinfo {pages}
  {037701} (\bibinfo {year} {2020})}\BibitemShut {NoStop}%
\bibitem [{\citenamefont {Zhang}\ \emph {et~al.}(2020)\citenamefont {Zhang},
  \citenamefont {Li}, \citenamefont {Huang},\ and\ \citenamefont
  {Zhang}}]{Zhang2020}%
  \BibitemOpen
  \bibfield  {author} {\bibinfo {author} {\bibfnamefont {J.-L.}\ \bibnamefont
  {Zhang}}, \bibinfo {author} {\bibfnamefont {Y.}~\bibnamefont {Li}}, \bibinfo
  {author} {\bibfnamefont {W.}~\bibnamefont {Huang}}, \ and\ \bibinfo {author}
  {\bibfnamefont {F.-C.}\ \bibnamefont {Zhang}},\ } {\bibfield  {journal} {\bibinfo  {journal}
  {Phys. Rev. B}\ }\textbf {\bibinfo {volume} {102}},\ \bibinfo {pages}
  {180509(R)} (\bibinfo {year} {2020})}\BibitemShut {NoStop}%
\bibitem [{\citenamefont {Guan}\ \emph {et~al.}(2022)\citenamefont {Guan},
  \citenamefont {Xiong}, \citenamefont {Wang},\ and\ \citenamefont
  {Luo}}]{Guan2022}%
  \BibitemOpen
  \bibfield  {author} {\bibinfo {author} {\bibfnamefont {S.}~\bibnamefont
  {Guan}}, \bibinfo {author} {\bibfnamefont {J.-X.}\ \bibnamefont {Xiong}},
  \bibinfo {author} {\bibfnamefont {Z.}~\bibnamefont {Wang}}, \ and\ \bibinfo
  {author} {\bibfnamefont {J.-W.}\ \bibnamefont {Luo}},\ } {\bibfield  {journal} {\bibinfo
  {journal} {Science China Physics, Mechanics \& Astronomy}\ }\textbf {\bibinfo
  {volume} {65}},\ \bibinfo {pages} {237301} (\bibinfo {year}
  {2022})}\BibitemShut {NoStop}%
\bibitem [{\citenamefont {Guan}\ and\ \citenamefont {Luo}(2020)}]{Guan2020a}%
  \BibitemOpen
  \bibfield  {author} {\bibinfo {author} {\bibfnamefont {S.}~\bibnamefont
  {Guan}}\ and\ \bibinfo {author} {\bibfnamefont {J.-W.}\ \bibnamefont {Luo}},\
  } {\bibfield  {journal} {\bibinfo
   {journal} {Phys. Rev. B}\ }\textbf {\bibinfo {volume} {102}},\ \bibinfo
  {pages} {184104} (\bibinfo {year} {2020})}\BibitemShut {NoStop}%
\bibitem [{\citenamefont {Dresselhaus}(1955)}]{Dresselhaus}%
  \BibitemOpen
  \bibfield  {author} {\bibinfo {author} {\bibfnamefont {G.}~\bibnamefont
  {Dresselhaus}},\ }\href@noop {} {\bibfield  {journal} {\bibinfo  {journal}
  {Phys. Rev.}\ }\textbf {\bibinfo {volume} {100}},\ \bibinfo {pages} {580}
  (\bibinfo {year} {1955})}\BibitemShut {NoStop}%
\bibitem [{\citenamefont {Bychkov}\ and\ \citenamefont
  {Rashba}(1984)}]{Rashba1984}%
  \BibitemOpen
  \bibfield  {author} {\bibinfo {author} {\bibfnamefont {Y.~A.}\ \bibnamefont
  {Bychkov}}\ and\ \bibinfo {author} {\bibfnamefont {E.~I.}\ \bibnamefont
  {Rashba}},\ }\href@noop {} {\bibfield  {journal} {\bibinfo  {journal} {JETP
  Lett.}\ }\textbf {\bibinfo {volume} {39}},\ \bibinfo {pages} {78} (\bibinfo
  {year} {1984})}\BibitemShut {NoStop}%
\bibitem [{\citenamefont {Manchon}\ \emph {et~al.}(2015)\citenamefont
  {Manchon}, \citenamefont {Koo}, \citenamefont {Nitta}, \citenamefont
  {Frolov},\ and\ \citenamefont {Duine}}]{Manchon2015}%
  \BibitemOpen
  \bibfield  {author} {\bibinfo {author} {\bibfnamefont {A.}~\bibnamefont
  {Manchon}}, \bibinfo {author} {\bibfnamefont {H.~C.}\ \bibnamefont {Koo}},
  \bibinfo {author} {\bibfnamefont {J.}~\bibnamefont {Nitta}}, \bibinfo
  {author} {\bibfnamefont {S.~M.}\ \bibnamefont {Frolov}}, \ and\ \bibinfo
  {author} {\bibfnamefont {R.~A.}\ \bibnamefont {Duine}},\ } {\bibfield  {journal} {\bibinfo  {journal}
  {Nature Materials}\ }\textbf {\bibinfo {volume} {14}},\ \bibinfo {pages}
  {871} (\bibinfo {year} {2015})}\BibitemShut {NoStop}%
\bibitem [{\citenamefont {Winkler}(2003)}]{Winkler2003}%
  \BibitemOpen
  \bibfield  {author} {\bibinfo {author} {\bibfnamefont {R.}~\bibnamefont
  {Winkler}},\ }\href@noop {} {\emph {\bibinfo {title} {Spin-Orbit Coupling
  Effect in Two-Dimensional Electron and Hole Systems}}}\ (\bibinfo
  {publisher} {Springer},\ \bibinfo {year} {2003})\BibitemShut {NoStop}%
\bibitem [{\citenamefont {Tao}\ \emph {et~al.}(2017)\citenamefont {Tao},
  \citenamefont {Paudel}, \citenamefont {Kovalev},\ and\ \citenamefont
  {Tsymbal}}]{Tao2017}%
  \BibitemOpen
  \bibfield  {author} {\bibinfo {author} {\bibfnamefont {L.~L.}\ \bibnamefont
  {Tao}}, \bibinfo {author} {\bibfnamefont {T.~R.}\ \bibnamefont {Paudel}},
  \bibinfo {author} {\bibfnamefont {A.~A.}\ \bibnamefont {Kovalev}}, \ and\
  \bibinfo {author} {\bibfnamefont {E.~Y.}\ \bibnamefont {Tsymbal}},\ } {\bibfield  {journal} {\bibinfo
  {journal} {Phys. Rev. B}\ }\textbf {\bibinfo {volume} {95}},\ \bibinfo
  {pages} {245141} (\bibinfo {year} {2017})}\BibitemShut {NoStop}%
\bibitem [{\citenamefont {Hirsch}(1999)}]{Hirsch1999}%
  \BibitemOpen
  \bibfield  {author} {\bibinfo {author} {\bibfnamefont {J.~E.}\ \bibnamefont
  {Hirsch}},\ } {\bibfield
  {journal} {\bibinfo  {journal} {Phys. Rev. Lett.}\ }\textbf {\bibinfo
  {volume} {83}},\ \bibinfo {pages} {1834} (\bibinfo {year}
  {1999})}\BibitemShut {NoStop}%
\bibitem [{\citenamefont {Zhang}(2000)}]{Zhang2000}%
  \BibitemOpen
  \bibfield  {author} {\bibinfo {author} {\bibfnamefont {S.}~\bibnamefont
  {Zhang}},\ } {\bibfield  {journal}
  {\bibinfo  {journal} {Phys. Rev. Lett.}\ }\textbf {\bibinfo {volume} {85}},\
  \bibinfo {pages} {393} (\bibinfo {year} {2000})}\BibitemShut {NoStop}%
\bibitem [{\citenamefont {Ganichev}\ \emph {et~al.}(2002)\citenamefont
  {Ganichev}, \citenamefont {Ivchenko}, \citenamefont {Bel'kov}, \citenamefont
  {Tarasenko}, \citenamefont {Sollinger}, \citenamefont {Weiss}, \citenamefont
  {Wegscheider},\ and\ \citenamefont {Prettl}}]{Ganichev2002}%
  \BibitemOpen
  \bibfield  {author} {\bibinfo {author} {\bibfnamefont {S.~D.}\ \bibnamefont
  {Ganichev}}, \bibinfo {author} {\bibfnamefont {E.~L.}\ \bibnamefont
  {Ivchenko}}, \bibinfo {author} {\bibfnamefont {V.~V.}\ \bibnamefont
  {Bel'kov}}, \bibinfo {author} {\bibfnamefont {S.~A.}\ \bibnamefont
  {Tarasenko}}, \bibinfo {author} {\bibfnamefont {M.}~\bibnamefont
  {Sollinger}}, \bibinfo {author} {\bibfnamefont {D.}~\bibnamefont {Weiss}},
  \bibinfo {author} {\bibfnamefont {W.}~\bibnamefont {Wegscheider}}, \ and\
  \bibinfo {author} {\bibfnamefont {W.}~\bibnamefont {Prettl}},\ } {\bibfield  {journal} {\bibinfo  {journal}
  {Nature}\ }\textbf {\bibinfo {volume} {417}},\ \bibinfo {pages} {153}
  (\bibinfo {year} {2002})}\BibitemShut {NoStop}%
\bibitem [{\citenamefont {Edelstein}(1990)}]{Edelstein1990}%
  \BibitemOpen
  \bibfield  {author} {\bibinfo {author} {\bibfnamefont {V.}~\bibnamefont
  {Edelstein}},\ }
  {\bibfield  {journal} {\bibinfo  {journal} {Solid State Communications}\
  }\textbf {\bibinfo {volume} {73}},\ \bibinfo {pages} {233} (\bibinfo {year}
  {1990})}\BibitemShut {NoStop}%
\bibitem [{\citenamefont {Datta}\ and\ \citenamefont {Das}(1990)}]{Datta1990}%
  \BibitemOpen
  \bibfield  {author} {\bibinfo {author} {\bibfnamefont {S.}~\bibnamefont
  {Datta}}\ and\ \bibinfo {author} {\bibfnamefont {B.}~\bibnamefont {Das}},\
  } {\bibfield  {journal} {\bibinfo  {journal}
  {Applied Physics Letters}\ }\textbf {\bibinfo {volume} {56}},\ \bibinfo
  {pages} {665} (\bibinfo {year} {1990})}, \BibitemShut {NoStop}%
\bibitem [{\citenamefont {Schliemann}(2017)}]{Schliemann2017a}%
  \BibitemOpen
  \bibfield  {author} {\bibinfo {author} {\bibfnamefont {J.}~\bibnamefont
  {Schliemann}},\ } {\bibfield
  {journal} {\bibinfo  {journal} {Rev. Mod. Phys.}\ }\textbf {\bibinfo {volume}
  {89}},\ \bibinfo {pages} {011001} (\bibinfo {year} {2017})}\BibitemShut
  {NoStop}%
\bibitem [{\citenamefont {Tao}\ and\ \citenamefont {Tsymbal}(2018)}]{Tao2018}%
  \BibitemOpen
  \bibfield  {author} {\bibinfo {author} {\bibfnamefont {L.~L.}\ \bibnamefont
  {Tao}}\ and\ \bibinfo {author} {\bibfnamefont {E.~Y.}\ \bibnamefont
  {Tsymbal}},\ } {\bibfield
  {journal} {\bibinfo  {journal} {Nature Communications}\ }\textbf {\bibinfo
  {volume} {9}},\ \bibinfo {pages} {2763} (\bibinfo {year} {2018})}\BibitemShut
  {NoStop}%
\bibitem [{\citenamefont {Bernevig}\ \emph {et~al.}(2006)\citenamefont
  {Bernevig}, \citenamefont {Orenstein},\ and\ \citenamefont
  {Zhang}}]{Bernevig2006}%
  \BibitemOpen
  \bibfield  {author} {\bibinfo {author} {\bibfnamefont {B.~A.}\ \bibnamefont
  {Bernevig}}, \bibinfo {author} {\bibfnamefont {J.}~\bibnamefont {Orenstein}},
  \ and\ \bibinfo {author} {\bibfnamefont {S.-C.}\ \bibnamefont {Zhang}},\
  } {\bibfield  {journal}
  {\bibinfo  {journal} {Phys. Rev. Lett.}\ }\textbf {\bibinfo {volume} {97}},\
  \bibinfo {pages} {236601} (\bibinfo {year} {2006})}\BibitemShut {NoStop}%
\bibitem [{\citenamefont {Kammermeier}\ \emph {et~al.}(2016)\citenamefont
  {Kammermeier}, \citenamefont {Wenk},\ and\ \citenamefont
  {Schliemann}}]{Kammermeier2016}%
  \BibitemOpen
  \bibfield  {author} {\bibinfo {author} {\bibfnamefont {M.}~\bibnamefont
  {Kammermeier}}, \bibinfo {author} {\bibfnamefont {P.}~\bibnamefont {Wenk}}, \
  and\ \bibinfo {author} {\bibfnamefont {J.}~\bibnamefont {Schliemann}},\
  } {\bibfield  {journal}
  {\bibinfo  {journal} {Phys. Rev. Lett.}\ }\textbf {\bibinfo {volume} {117}},\
  \bibinfo {pages} {236801} (\bibinfo {year} {2016})}\BibitemShut {NoStop}%
\bibitem [{\citenamefont {Koralek}\ \emph {et~al.}(2009)\citenamefont
  {Koralek}, \citenamefont {Weber}, \citenamefont {Orenstein}, \citenamefont
  {Bernevig}, \citenamefont {Zhang}, \citenamefont {Mack},\ and\ \citenamefont
  {Awschalom}}]{Koralek2009}%
  \BibitemOpen
  \bibfield  {author} {\bibinfo {author} {\bibfnamefont {J.~D.}\ \bibnamefont
  {Koralek}}, \bibinfo {author} {\bibfnamefont {C.~P.}\ \bibnamefont {Weber}},
  \bibinfo {author} {\bibfnamefont {J.}~\bibnamefont {Orenstein}}, \bibinfo
  {author} {\bibfnamefont {B.~A.}\ \bibnamefont {Bernevig}}, \bibinfo {author}
  {\bibfnamefont {S.-C.}\ \bibnamefont {Zhang}}, \bibinfo {author}
  {\bibfnamefont {S.}~\bibnamefont {Mack}}, \ and\ \bibinfo {author}
  {\bibfnamefont {D.~D.}\ \bibnamefont {Awschalom}},\ } {\bibfield  {journal} {\bibinfo
  {journal} {Nature}\ }\textbf {\bibinfo {volume} {458}},\ \bibinfo {pages}
  {610} (\bibinfo {year} {2009})}\BibitemShut {NoStop}%
\bibitem [{\citenamefont {Walser}\ \emph {et~al.}(2012)\citenamefont {Walser},
  \citenamefont {Reichl}, \citenamefont {Wegscheider},\ and\ \citenamefont
  {Salis}}]{Walser2012}%
  \BibitemOpen
  \bibfield  {author} {\bibinfo {author} {\bibfnamefont {M.~P.}\ \bibnamefont
  {Walser}}, \bibinfo {author} {\bibfnamefont {C.}~\bibnamefont {Reichl}},
  \bibinfo {author} {\bibfnamefont {W.}~\bibnamefont {Wegscheider}}, \ and\
  \bibinfo {author} {\bibfnamefont {G.}~\bibnamefont {Salis}},\ } {\bibfield  {journal} {\bibinfo
  {journal} {Nature Physics}\ }\textbf {\bibinfo {volume} {8}},\ \bibinfo
  {pages} {757} (\bibinfo {year} {2012})}\BibitemShut {NoStop}%
\bibitem [{\citenamefont {Kohda}\ \emph {et~al.}(2012)\citenamefont {Kohda},
  \citenamefont {Lechner}, \citenamefont {Kunihashi}, \citenamefont
  {Dollinger}, \citenamefont {Olbrich}, \citenamefont {Sch\"onhuber},
  \citenamefont {Caspers}, \citenamefont {Bel'kov}, \citenamefont {Golub},
  \citenamefont {Weiss}, \citenamefont {Richter}, \citenamefont {Nitta},\ and\
  \citenamefont {Ganichev}}]{Kohda2012}%
  \BibitemOpen
  \bibfield  {author} {\bibinfo {author} {\bibfnamefont {M.}~\bibnamefont
  {Kohda}}, \bibinfo {author} {\bibfnamefont {V.}~\bibnamefont {Lechner}},
  \bibinfo {author} {\bibfnamefont {Y.}~\bibnamefont {Kunihashi}}, \bibinfo
  {author} {\bibfnamefont {T.}~\bibnamefont {Dollinger}}, \bibinfo {author}
  {\bibfnamefont {P.}~\bibnamefont {Olbrich}}, \bibinfo {author} {\bibfnamefont
  {C.}~\bibnamefont {Sch\"onhuber}}, \bibinfo {author} {\bibfnamefont
  {I.}~\bibnamefont {Caspers}}, \bibinfo {author} {\bibfnamefont {V.~V.}\
  \bibnamefont {Bel'kov}}, \bibinfo {author} {\bibfnamefont {L.~E.}\
  \bibnamefont {Golub}}, \bibinfo {author} {\bibfnamefont {D.}~\bibnamefont
  {Weiss}}, \bibinfo {author} {\bibfnamefont {K.}~\bibnamefont {Richter}},
  \bibinfo {author} {\bibfnamefont {J.}~\bibnamefont {Nitta}}, \ and\ \bibinfo
  {author} {\bibfnamefont {S.~D.}\ \bibnamefont {Ganichev}},\ } {\bibfield  {journal} {\bibinfo  {journal} {Phys.
  Rev. B}\ }\textbf {\bibinfo {volume} {86}},\ \bibinfo {pages} {081306(R)}
  (\bibinfo {year} {2012})}\BibitemShut {NoStop}%
\bibitem [{\citenamefont {Sasaki}\ \emph {et~al.}(2014)\citenamefont {Sasaki},
  \citenamefont {Nonaka}, \citenamefont {Kunihashi}, \citenamefont {Kohda},
  \citenamefont {Bauernfeind}, \citenamefont {Dollinger}, \citenamefont
  {Richter},\ and\ \citenamefont {Nitta}}]{Sasaki2014}%
  \BibitemOpen
  \bibfield  {author} {\bibinfo {author} {\bibfnamefont {A.}~\bibnamefont
  {Sasaki}}, \bibinfo {author} {\bibfnamefont {S.}~\bibnamefont {Nonaka}},
  \bibinfo {author} {\bibfnamefont {Y.}~\bibnamefont {Kunihashi}}, \bibinfo
  {author} {\bibfnamefont {M.}~\bibnamefont {Kohda}}, \bibinfo {author}
  {\bibfnamefont {T.}~\bibnamefont {Bauernfeind}}, \bibinfo {author}
  {\bibfnamefont {T.}~\bibnamefont {Dollinger}}, \bibinfo {author}
  {\bibfnamefont {K.}~\bibnamefont {Richter}}, \ and\ \bibinfo {author}
  {\bibfnamefont {J.}~\bibnamefont {Nitta}},\ } {\bibfield  {journal} {\bibinfo
  {journal} {Nature Nanotechnology}\ }\textbf {\bibinfo {volume} {9}},\
  \bibinfo {pages} {703} (\bibinfo {year} {2014})}\BibitemShut {NoStop}%
\bibitem [{\citenamefont {Ji}\ \emph {et~al.}(2022)\citenamefont {Ji},
  \citenamefont {Lou}, \citenamefont {Yu}, \citenamefont {Feng},\ and\
  \citenamefont {Xiang}}]{Ji2022}%
  \BibitemOpen
  \bibfield  {author} {\bibinfo {author} {\bibfnamefont {J.}~\bibnamefont
  {Ji}}, \bibinfo {author} {\bibfnamefont {F.}~\bibnamefont {Lou}}, \bibinfo
  {author} {\bibfnamefont {R.}~\bibnamefont {Yu}}, \bibinfo {author}
  {\bibfnamefont {J.~S.}\ \bibnamefont {Feng}}, \ and\ \bibinfo {author}
  {\bibfnamefont {H.~J.}\ \bibnamefont {Xiang}},\ } {\bibfield  {journal} {\bibinfo  {journal}
  {Phys. Rev. B}\ }\textbf {\bibinfo {volume} {105}},\ \bibinfo {pages}
  {L041404} (\bibinfo {year} {2022})}\BibitemShut {NoStop}%
\bibitem [{\citenamefont {See the Supplemental Material at http://link.aps.org/
  for details of the first-principles calculation}()}]{sp}%
  \BibitemOpen
  \bibfield  {author} {\bibinfo {author} { \bibnamefont {See Supplemental Material for the 
  first-principle calculations details, hidden Zeeman-type spin polarization for WSe$_2$ 
  at k$_z$=$\pi$, and band structure without considering SOC for BaBi$_4$O$_7$, which includes Refs.~\cite{Kresse1993,Kresse1996,Bloechl1994,PBE,Dion2004}}}\
  }\href@noop {} {\ }\BibitemShut {NoStop}%
\bibitem[{\citenamefont{Kresse and Hafner}(1993)}]{Kresse1993}
\bibinfo{author}{\bibfnamefont{G.}~\bibnamefont{Kresse}} \bibnamefont{and}
  \bibinfo{author}{\bibfnamefont{J.}~\bibnamefont{Hafner}},
  \bibinfo{journal}{Phys. Rev. B} \textbf{\bibinfo{volume}{47}},
  \bibinfo{pages}{558} (\bibinfo{year}{1993}).

\bibitem[{\citenamefont{Kresse and Furthm\"uller}(1996)}]{Kresse1996}
\bibinfo{author}{\bibfnamefont{G.}~\bibnamefont{Kresse}} \bibnamefont{and}
  \bibinfo{author}{\bibfnamefont{J.}~\bibnamefont{Furthm\"uller}},
  \bibinfo{journal}{Phys. Rev. B} \textbf{\bibinfo{volume}{54}},
  \bibinfo{pages}{11169} (\bibinfo{year}{1996}).

\bibitem[{\citenamefont{Bl\"ochl}(1994)}]{Bloechl1994}
\bibinfo{author}{\bibfnamefont{P.~E.} \bibnamefont{Bl\"ochl}},
  \bibinfo{journal}{Phys. Rev. B} \textbf{\bibinfo{volume}{50}},
  \bibinfo{pages}{17953} (\bibinfo{year}{1994}).

\bibitem[{\citenamefont{Perdew et~al.}(1996)\citenamefont{Perdew, Burke, and
  Ernzerhof}}]{PBE}
\bibinfo{author}{\bibfnamefont{J.~P.} \bibnamefont{Perdew}},
  \bibinfo{author}{\bibfnamefont{K.}~\bibnamefont{Burke}}, \bibnamefont{and}
  \bibinfo{author}{\bibfnamefont{M.}~\bibnamefont{Ernzerhof}},
  \bibinfo{journal}{Phys. Rev. Lett.} \textbf{\bibinfo{volume}{77}},
  \bibinfo{pages}{3865} (\bibinfo{year}{1996}).

\bibitem[{\citenamefont{Dion et~al.}(2004)\citenamefont{Dion, Rydberg,
  Schr\"oder, Langreth, and Lundqvist}}]{Dion2004}
\bibinfo{author}{\bibfnamefont{M.}~\bibnamefont{Dion}},
  \bibinfo{author}{\bibfnamefont{H.}~\bibnamefont{Rydberg}},
  \bibinfo{author}{\bibfnamefont{E.}~\bibnamefont{Schr\"oder}},
  \bibinfo{author}{\bibfnamefont{D.~C.} \bibnamefont{Langreth}},
  \bibnamefont{and} \bibinfo{author}{\bibfnamefont{B.~I.}
  \bibnamefont{Lundqvist}}, \bibinfo{journal}{Phys. Rev. Lett.}
  \textbf{\bibinfo{volume}{92}}, \bibinfo{pages}{246401}
  (\bibinfo{year}{2004}).

\bibitem [{\citenamefont {Coehoorn}\ \emph {et~al.}(1987)\citenamefont
  {Coehoorn}, \citenamefont {Haas}, \citenamefont {Dijkstra}, \citenamefont
  {Flipse}, \citenamefont {de~Groot},\ and\ \citenamefont
  {Wold}}]{Coehoorn1987}%
  \BibitemOpen
  \bibfield  {author} {\bibinfo {author} {\bibfnamefont {R.}~\bibnamefont
  {Coehoorn}}, \bibinfo {author} {\bibfnamefont {C.}~\bibnamefont {Haas}},
  \bibinfo {author} {\bibfnamefont {J.}~\bibnamefont {Dijkstra}}, \bibinfo
  {author} {\bibfnamefont {C.~J.~F.}\ \bibnamefont {Flipse}}, \bibinfo {author}
  {\bibfnamefont {R.~A.}\ \bibnamefont {de~Groot}}, \ and\ \bibinfo {author}
  {\bibfnamefont {A.}~\bibnamefont {Wold}},\ } {\bibfield  {journal} {\bibinfo  {journal} {Phys.
  Rev. B}\ }\textbf {\bibinfo {volume} {35}},\ \bibinfo {pages} {6195}
  (\bibinfo {year} {1987})}\BibitemShut {NoStop}%
\bibitem [{\citenamefont {Zhong}\ \emph {et~al.}(2016)\citenamefont {Zhong},
  \citenamefont {Chen}, \citenamefont {Xie}, \citenamefont {Yang},
  \citenamefont {Cohen},\ and\ \citenamefont {Zhang}}]{Zhong2016}%
  \BibitemOpen
  \bibfield  {author} {\bibinfo {author} {\bibfnamefont {C.}~\bibnamefont
  {Zhong}}, \bibinfo {author} {\bibfnamefont {Y.}~\bibnamefont {Chen}},
  \bibinfo {author} {\bibfnamefont {Y.}~\bibnamefont {Xie}}, \bibinfo {author}
  {\bibfnamefont {S.~A.}\ \bibnamefont {Yang}}, \bibinfo {author}
  {\bibfnamefont {M.~L.}\ \bibnamefont {Cohen}}, \ and\ \bibinfo {author}
  {\bibfnamefont {S.~B.}\ \bibnamefont {Zhang}},\ } {\bibfield  {journal} {\bibinfo  {journal} {Nanoscale}\
  }\textbf {\bibinfo {volume} {8}},\ \bibinfo {pages} {7232} (\bibinfo {year}
  {2016})}\BibitemShut {NoStop}%
\bibitem [{\citenamefont {Liang}\ \emph {et~al.}(2016)\citenamefont {Liang},
  \citenamefont {Zhou}, \citenamefont {Yu}, \citenamefont {Wang},\ and\
  \citenamefont {Weng}}]{Liang2016}%
  \BibitemOpen
  \bibfield  {author} {\bibinfo {author} {\bibfnamefont {Q.-F.}\ \bibnamefont
  {Liang}}, \bibinfo {author} {\bibfnamefont {J.}~\bibnamefont {Zhou}},
  \bibinfo {author} {\bibfnamefont {R.}~\bibnamefont {Yu}}, \bibinfo {author}
  {\bibfnamefont {Z.}~\bibnamefont {Wang}}, \ and\ \bibinfo {author}
  {\bibfnamefont {H.}~\bibnamefont {Weng}},\ } {\bibfield  {journal} {\bibinfo  {journal} {Phys.
  Rev. B}\ }\textbf {\bibinfo {volume} {93}},\ \bibinfo {pages} {085427}
  (\bibinfo {year} {2016})}\BibitemShut {NoStop}%
\bibitem [{\citenamefont {Wu}\ \emph {et~al.}(2018)\citenamefont {Wu},
  \citenamefont {Liu}, \citenamefont {Li}, \citenamefont {Zhong}, \citenamefont
  {Yu}, \citenamefont {Sheng}, \citenamefont {Zhao},\ and\ \citenamefont
  {Yang}}]{Wu2018}%
  \BibitemOpen
  \bibfield  {author} {\bibinfo {author} {\bibfnamefont {W.}~\bibnamefont
  {Wu}}, \bibinfo {author} {\bibfnamefont {Y.}~\bibnamefont {Liu}}, \bibinfo
  {author} {\bibfnamefont {S.}~\bibnamefont {Li}}, \bibinfo {author}
  {\bibfnamefont {C.}~\bibnamefont {Zhong}}, \bibinfo {author} {\bibfnamefont
  {Z.-M.}\ \bibnamefont {Yu}}, \bibinfo {author} {\bibfnamefont {X.-L.}\
  \bibnamefont {Sheng}}, \bibinfo {author} {\bibfnamefont {Y.~X.}\ \bibnamefont
  {Zhao}}, \ and\ \bibinfo {author} {\bibfnamefont {S.~A.}\ \bibnamefont
  {Yang}},\ } {\bibfield  {journal}
  {\bibinfo  {journal} {Phys. Rev. B}\ }\textbf {\bibinfo {volume} {97}},\
  \bibinfo {pages} {115125} (\bibinfo {year} {2018})}\BibitemShut {NoStop}%
\bibitem [{\citenamefont {Klinkova}\ \emph {et~al.}(1996)\citenamefont
  {Klinkova}, \citenamefont {Nikolaichik}, \citenamefont {Zorina},
  \citenamefont {Barkovskii}, \citenamefont {Fedotov},\ and\ \citenamefont
  {Zver'Kov}}]{Klinkova1996}%
  \BibitemOpen
  \bibfield  {author} {\bibinfo {author} {\bibfnamefont {L.}~\bibnamefont
  {Klinkova}}, \bibinfo {author} {\bibfnamefont {V.}~\bibnamefont
  {Nikolaichik}}, \bibinfo {author} {\bibfnamefont {L.}~\bibnamefont {Zorina}},
  \bibinfo {author} {\bibfnamefont {N.}~\bibnamefont {Barkovskii}}, \bibinfo
  {author} {\bibfnamefont {V.}~\bibnamefont {Fedotov}}, \ and\ \bibinfo
  {author} {\bibfnamefont {A.}~\bibnamefont {Zver'Kov}},\ }\href@noop {}
  {\bibfield  {journal} {\bibinfo  {journal} {Zhurnal Neorganicheskoj Khimii}\
  }\textbf {\bibinfo {volume} {41}},\ \bibinfo {pages} {709} (\bibinfo {year}
  {1996})}\BibitemShut {NoStop}%
\bibitem [{\citenamefont {Bradley}\ and\ \citenamefont
  {Cracknell}(1972)}]{Bradley1972}%
  \BibitemOpen
  \bibfield  {author} {\bibinfo {author} {\bibfnamefont {C.~J.}\ \bibnamefont
  {Bradley}}\ and\ \bibinfo {author} {\bibfnamefont {A.~P.}\ \bibnamefont
  {Cracknell}},\ }\href@noop {} {\emph {\bibinfo {title} {The Mathematical
  Theory of Symmetry in Solids}}}\ (\bibinfo  {publisher} {Clarendon, Oxford},\
  \bibinfo {year} {1972})\BibitemShut {NoStop}%
\bibitem [{\citenamefont {Aroyo}\ \emph {et~al.}(2006)\citenamefont {Aroyo},
  \citenamefont {Perez-Mato}, \citenamefont {Capillas}, \citenamefont
  {Kroumova}, \citenamefont {Ivantchev}, \citenamefont {Madariaga},
  \citenamefont {Kirov},\ and\ \citenamefont {Wondratschek}}]{Aroyo2006}%
  \BibitemOpen
  \bibfield  {author} {\bibinfo {author} {\bibfnamefont {M.~I.}\ \bibnamefont
  {Aroyo}}, \bibinfo {author} {\bibfnamefont {J.~M.}\ \bibnamefont
  {Perez-Mato}}, \bibinfo {author} {\bibfnamefont {C.}~\bibnamefont
  {Capillas}}, \bibinfo {author} {\bibfnamefont {E.}~\bibnamefont {Kroumova}},
  \bibinfo {author} {\bibfnamefont {S.}~\bibnamefont {Ivantchev}}, \bibinfo
  {author} {\bibfnamefont {G.}~\bibnamefont {Madariaga}}, \bibinfo {author}
  {\bibfnamefont {A.}~\bibnamefont {Kirov}}, \ and\ \bibinfo {author}
  {\bibfnamefont {H.}~\bibnamefont {Wondratschek}},\ } {\bibfield  {journal} {\bibinfo  {journal}
  {Zeitschrift fr Kristallographie - Crystalline Materials}\ }\textbf
  {\bibinfo {volume} {221}},\ \bibinfo {pages} {15} (\bibinfo {year}
  {2006})}\BibitemShut {NoStop}%
\bibitem [{\citenamefont {Ram\'{\i}rez-Ruiz}\ \emph {et~al.}(2017)\citenamefont
  {Ram\'{\i}rez-Ruiz}, \citenamefont {Boutin},\ and\ \citenamefont
  {Garate}}]{Ramirez-Ruiz2017}%
  \BibitemOpen
  \bibfield  {author} {\bibinfo {author} {\bibfnamefont {J.}~\bibnamefont
  {Ram\'{\i}rez-Ruiz}}, \bibinfo {author} {\bibfnamefont {S.}~\bibnamefont
  {Boutin}}, \ and\ \bibinfo {author} {\bibfnamefont {I.}~\bibnamefont
  {Garate}},\ } {\bibfield
  {journal} {\bibinfo  {journal} {Phys. Rev. B}\ }\textbf {\bibinfo {volume}
  {96}},\ \bibinfo {pages} {235201} (\bibinfo {year} {2017})}\BibitemShut
  {NoStop}%
\bibitem [{\citenamefont {Gehlmann}\ \emph {et~al.}(2016)\citenamefont
  {Gehlmann}, \citenamefont {Aguilera}, \citenamefont {Bihlmayer},
  \citenamefont {Mlynczak}, \citenamefont {Eschbach}, \citenamefont
  {Doring}, \citenamefont {Gospodaric}, \citenamefont {Cramm},
  \citenamefont {Kardynal}, \citenamefont {Plucinski}, \citenamefont
  {Blugel},\ and\ \citenamefont {Schneider}}]{Gehlmann2016}%
  \BibitemOpen
  \bibfield  {author} {\bibinfo {author} {\bibfnamefont {M.}~\bibnamefont
  {Gehlmann}}, \bibinfo {author} {\bibfnamefont {I.}~\bibnamefont {Aguilera}},
  \bibinfo {author} {\bibfnamefont {G.}~\bibnamefont {Bihlmayer}}, \bibinfo
  {author} {\bibfnamefont {E.}~\bibnamefont {Mlynczak}}, \bibinfo {author}
  {\bibfnamefont {M.}~\bibnamefont {Eschbach}}, \bibinfo {author}
  {\bibfnamefont {S.}~\bibnamefont {Doring}}, \bibinfo {author}
  {\bibfnamefont {P.}~\bibnamefont {Gospodaric}}, \bibinfo {author}
  {\bibfnamefont {S.}~\bibnamefont {Cramm}}, \bibinfo {author} {\bibfnamefont
  {B.}~\bibnamefont {Kardynal}}, \bibinfo {author} {\bibfnamefont
  {L.}~\bibnamefont {Plucinski}}, \bibinfo {author} {\bibfnamefont
  {S.}~\bibnamefont {Blugel}}, \ and\ \bibinfo {author} {\bibfnamefont
  {C.~M.}\ \bibnamefont {Schneider}},\ } {\bibfield  {journal} {\bibinfo
  {journal} {Scientific Reports}\ }\textbf {\bibinfo {volume} {6}},\ \bibinfo
  {pages} {26197} (\bibinfo {year} {2016})}\BibitemShut {NoStop}%
\bibitem [{\citenamefont {Li}\ and\ \citenamefont {Appelbaum}(2018)}]{Li2018}%
  \BibitemOpen
  \bibfield  {author} {\bibinfo {author} {\bibfnamefont {P.}~\bibnamefont
  {Li}}\ and\ \bibinfo {author} {\bibfnamefont {I.}~\bibnamefont {Appelbaum}},\
  } {\bibfield  {journal} {\bibinfo
  {journal} {Phys. Rev. B}\ }\textbf {\bibinfo {volume} {97}},\ \bibinfo
  {pages} {125434} (\bibinfo {year} {2018})}\BibitemShut {NoStop}%

\end{thebibliography}
%

\end{document}